\documentclass[preprint]{aastex}
\usepackage{emulateapj5,psfig}
\usepackage{apjfonts}\tighten
\usepackage{graphicx}

\newcommand{\as}{\mbox{$''$}}
\newcommand{\apsis}{{\sl Apsis\/}}

\newcommand{\ebv}{\mbox{$E(B-V)$}}
\newcommand{\etagc}{\mbox{$\eta_{\rm GC}$}}
\newcommand{\etal}{{\it et~al.\/}}
\newcommand{\epscl}{\mbox{$\epsilon_{\rm cl}$}}

\newcommand{\feh}{\mbox{[Fe/H]}}
\newcommand{\HI}{\mbox{\sc H\,{i}}}

\newcommand{\MHILB}{\mbox{${\cal M}_{\rm HI}/L_B$}}
\newcommand{\MTLB}{\mbox{${\cal M}_T/L_B$}}

\newcommand{\Msun}{\mbox{${\cal M}_\odot$}}
\newcommand{\gacs}{\mbox{$g_{475}$}}
\newcommand{\vacs}{\mbox{$V_{606}$}}
\newcommand{\iacs}{\mbox{$I_{814}$}}
\newcommand{\Iacs}{\mbox{$I_{814,0}$}}

\newcommand{\GVacs}{\mbox{$(g_{475} - V_{606})_0$}}
\newcommand{\VIacs}{\mbox{$(V_{606} - I_{814})_0$}}

\newcommand{\rhalf}{\mbox{$r_{1/2}$}}
\newcommand{\SN}{\mbox{$S_N$}}

\shorttitle{Globular clusters in NGC\,2915}
\shortauthors{Meurer et al.}

\submitted{ApJ Letters, accepted}
\begin{document}


\title{Discovery of Globular Clusters in the Proto-Spiral 
  NGC\,2915: Implications for Hierarchical Galaxy
  Evolution}


\author{Gerhardt R.\ Meurer\altaffilmark{1}, J.P.\
  Blakeslee\altaffilmark{1}, M.\ Sirianni\altaffilmark{1}, H.C.\
  Ford\altaffilmark{1}, G.D. Illingworth\altaffilmark{2},
  N. Ben\'{\i}tez\altaffilmark{1}, M. Clampin\altaffilmark{3},
  F. Menanteau\altaffilmark{1}, H.D. Tran\altaffilmark{4}, 
  R.A. Kimble\altaffilmark{3}, G.F. Hartig\altaffilmark{5},
  D.R. Ardila\altaffilmark{1}, F. Bartko\altaffilmark{6},
  R.J. Bouwens\altaffilmark{1}, T.J. Broadhurst\altaffilmark{7},
  R.A. Brown\altaffilmark{5}, C.J. Burrows\altaffilmark{5},
  E.S. Cheng\altaffilmark{3}, N.J.G. Cross\altaffilmark{1},
  P.D. Feldman\altaffilmark{1}, D.A. Golimowski\altaffilmark{1},
  C. Gronwall\altaffilmark{8}, L. Infante\altaffilmark{9},
  J.E. Krist\altaffilmark{5}, M.P. Lesser\altaffilmark{10},
  A.R. Martel\altaffilmark{1}, 
  G.K. Miley\altaffilmark{11}, M. Postman\altaffilmark{5},
  P. Rosati\altaffilmark{12}, W.B. Sparks\altaffilmark{5},
  Z.I. Tsvetanov\altaffilmark{1},
  R.L. White\altaffilmark{1,5}, \& W. Zheng\altaffilmark{1}}
  
\altaffiltext{1}{Department of Physics and Astronomy, The Johns Hopkins
University, 3400 North Charles Street, Baltimore, MD 21218; meurer@pha.jhu.edu}

\altaffiltext{2}{UCO/Lick Observatory, University of California, Santa
Cruz, CA 95064.}


\altaffiltext{3}{NASA Goddard Space Flight Center, Greenbelt, MD 20771.}


\altaffiltext{4}{W.M.\ Keck Observatory, 65-1120 Mamalahoa Hwy., Kamuela, HI 96743}

\altaffiltext{5}{STScI, 3700 San Martin Drive, Baltimore, MD 21218.}


\altaffiltext{6}{Bartko Science \& Technology, P.O. Box 670, Mead, CO
80542-0670.}


\altaffiltext{7}{Racah Institute of Physics, The Hebrew University,
Jerusalem, Israel 91904.}


\altaffiltext{8}{Department of Astronomy and Astrophysics, The
Pennsylvania State University, 525 Davey Lab, University Park, PA
16802.}


\altaffiltext{9}{Departmento de Astronom\'{\i}a y Astrof\'{\i}sica,
Pontificia Universidad Cat\'{\o}lica de Chile, Casilla 306, Santiago
22, Chile.}


\altaffiltext{10}{Steward Observatory, University of Arizona, Tucson,
AZ 85721.}


\altaffiltext{11}{Leiden Observatory, Postbus 9513, 2300 RA Leiden,
Netherlands.}


\altaffiltext{12}{European Southern Observatory,
Karl-Schwarzschild-Strasse 2, D-85748 Garching, Germany.}

%


\begin{abstract}
We have discovered three globular clusters beyond the Holmberg radius in
Hubble Space Telescope Advanced Camera for Surveys images of the
gas-rich dark matter dominated blue compact dwarf galaxy NGC\,2915.  The
clusters, all of which start to resolve into stars, have $M_{V606} =
-8.9$ to --9.8 mag, significantly brighter than the peak of the
luminosity function of Milky Way globular clusters. Their colors suggest
a metallicity $[{\rm Fe/H}] \approx -1.9$ dex, typical of metal-poor
Galactic globular clusters.  The specific frequency of clusters is at a
minimum normal, compared to spiral galaxies.  However, since only a
small portion of the system has been surveyed it is more likely that the
luminosity and mass normalized cluster content is higher, like that seen
in elliptical galaxies and galaxy clusters.  This suggests that NGC\,2915
resembles a key phase in the early hierarchical assembly of galaxies -
the epoch when much of the old stellar population has formed, but little
of the stellar disk.  Depending on the subsequent interaction history,
such systems could go on to build-up larger elliptical galaxies, evolve
into normal spirals, or in rare circumstances remain suspended in their
development to become systems like NGC\,2915.
\end{abstract}




\section{Introduction}\label{s:intro}

All galaxies with massive old stellar populations are thought to contain
globular clusters (GCs).  They are particularly noticeable where the old
stellar population is dominant such as in elliptical (E), dwarf
elliptical (dE), and central Dominant (cD), as well as spiral galaxies
with prominent bulges.  Dwarf spheroidal galaxies generally do not
contain GCs, probably because they have insufficient mass to make their
formation likely.  The exceptions are the most massive dwarf spheroidals
Fornax \citep{h61} and Sagitarius \citep{igi94} which each have at least
4 GCs.  Disk galaxies dominated by population I stars contain fewer GCs
per unit luminosity, presumably because of star formation in the disk
after the formation of the population II component.  Galaxies with a
high \MHILB\ ratio have yet to form much of their baryonic mass into
stars.  They typically are blue and not considered likely hosts for
populous GC systems.

NGC\,2915 is an extreme gas rich galaxy having $\MHILB = 1.7\,
\Msun/L_{B,\odot}$ \citep[][hereafter MCBF96]{mcbf96}.  Its regularly
rotating \HI\ disk extends to over 5 times beyond the readily
detectable optical emission providing an excellent dynamical tracer for
the mass distribution; not coincidentally NGC\,2915 has one of the
highest known mass-to-light ratios in a single galaxy (MCBF96).
Furthermore, while its optical morphology is that of a blue compact
dwarf \citep[BCD;][hereafter MMC94]{mmc94}, its \HI\ disk clearly shows
spiral arms which are not apparent in the optical.

In this Letter, we report the discovery of three luminous GCs found in
Hubble Space Telescope Advanced Camera for Surveys
\citep[ACS;][]{ford_acs02} images of NGC\,2915 that were obtained in
order to look for a stellar heating source for the \HI\ disk.  That
issue will be discussed in a separate article (Meurer \etal\ 2003; in
preparation, hereafter Meu03). 

\section{Data and analysis}\label{s:data}

ACS Wide Field Camera (WFC) images were obtained of a field centered at
09$^{\rm h}$ 25$^{\rm m}$ 36$\fs$48, --76$^\circ$ 35$'$ 52$\farcs$4
(J2000).  The images cover projected radii of 45\as\ to 257 \as, whereas
the Holmberg radius $R_{Ho} = 114''$. We obtained 2, 2, 4 images for a
total exposure of 2480$s$, 2600$s$, 5220$s$ in the filters F475W
(\gacs), F606W (\vacs), and F814W (\iacs), respectively.  The basic
processing of the images was done using the {\sl CALACS} pipeline
\citep{hack99}.  We used the program \apsis\ \citep{bambm02} to align
and combine the images encorporating geometric correction and rejection
of cosmic rays and hot pixels. 

Here we present photometry in the natural system of the filters, with
zeropoints selected so that Vega would have a magnitude of 0.0 in all
bands.  In order to compare our observations with previous work, we
convert the previous work to this system, as needed, using the
calibrations of Sirianni et al (2003, in preparation).  The most
important correction is to the \vacs\ photometry, since the F606W filter
straddles the wavelength of traditional $V$ and $R$ filters.

\section{Results}\label{s:res}

Table~\ref{t:prop} presents adopted global properties for NGC\,2915.  The
foreground extinction, \ebv, is from the \citet{sfd98} extinction maps.
It is significantly larger than $\ebv = 0.15 \pm 0.05$ estimated by
MMC94, but consistent with the position of the field star Red Giant
Branch (RGB; Meu03).  Extinction corrected photometry employing the
\citet{ccm89} extinction curve is denoted with a ``0'' subscript.  The
distance, $D$ was derived from the field star RGB tip (Meu03).  It is
consistent with but improves on previous estimates $D = 5.3 \pm 1.3$ Mpc
(MMC94) and $D = 3.8 \pm 0.5$ Mpc \citep{k03}.  The remaining quantities
in Table~\ref{t:prop} were derived from MMC94 and MCBF96 after
correcting to the new \ebv\ and $D$.

\begin{figure}
\epsscale{1.0}
\plotone{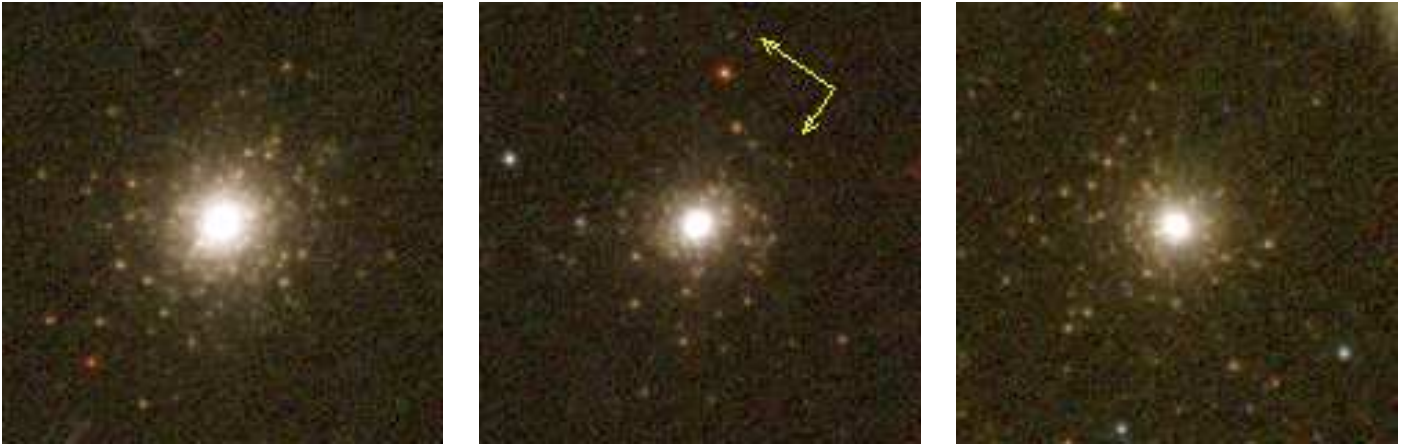}
\caption[]{Three color images of the three globular clusters; from left
to right: G1, G2, G3.  The images are shown on the same logarithmic
stretch, with F814W, F606W, and F475W representing red, green and blue
respectively.  Each image is 8\as\ on a side and has the same
orientation.  The large arrow of the compass in the middle panel points
to north, the small arrow points to east.
\label{f:finders}}
\end{figure}

As shown in Fig.~\ref{f:finders}, the three sources are clearly GCs
whose brightest stars are resolved. Table~\ref{t:clust} compiles the
properties of the clusters.  The photometric quantities were measured
using a circular aperture having a radius of $r = 3''$, with the local
sky subtracted using an annulus having radii of 5\as\ and 7.5\as.  The
cluster size \rhalf\ is the circular aperture radius encompassing half
the \vacs\ light as measured from curves of growth.

Compared to Galactic GCs, these clusters are large and luminous, but not
abnormally so.  Only 16\%\ of the clusters in the \citet{h96}
database\footnote{http://physwww.physics.mcmaster.ca/$\sim$harris/mwgc.dat}
have \vacs\ luminosities brighter than G3; only three clusters are
more luminous than G1.  The clusters' \rhalf\ ranges from about 5 to 9
pc, placing them in the upper quartile of Galactic GCs which have
\rhalf\ ranging from 0.3 to 24.7 pc \citep{h96}.  The clusters are
noticeably elongated with ellipticity $\epsilon \equiv 1 - b/a$ similar
to the canonical flattened Galactic GCs M22 and $\omega$ Cen ($\epsilon
= 0.14$, 0.17, respectively; Harris 1996).  The combination of high
luminosity and appreciable flattening is also seen in the cluster M31-G1
\citep{pvdb84,msjdbr01}.

The \GVacs\ and \VIacs\ colors of the clusters are compared to Milky Way
GCs \citep{h96} in Fig.~\ref{f:2cd}. Their colors are virtually
identical implying similar metallicities, assuming they are old and
nearly coeval.  We derive their metallicity by fitting the metallicty -
color relationship from the Harris database after converting the colors
to \VIacs.  We employed an unweighted least squares fit with an
iterative $2.5\sigma$ rejection resulting in $\feh = -5.37 + 5.36\VIacs$
with a dispersion of 0.29 dex.  The metallicity for the three clusters
is then $\feh = -1.9 \pm 0.4$ dex, consistent with {\em low\/}
metallicity Galactic GCs.  Our stellar population analysis, in progress
(Meu03), indicates that the stars at the outskirts of the clusters have
very similar \Iacs\ versus \VIacs\ color-magnitude diagrams, dominated
by a narrow and blue RGB.  This is also consistent with low \feh, if the
clusters are old.

\section{Discussion}\label{s:disc}

\subsection{Cluster formation efficiency}

Because of the blue core and gas rich nature of NGC\,2915 we had not
expected to find GCs in our images.  However, in retrospect this
discovery should not have been surprising. GCs have previously been
discovered around morphologically similar systems; \citet{obr98} find a
considerable population of at least 65 old GCs about the Blue Compact
(albeit not dwarf) galaxy ESO~338-IG04.  In addition, NGC\,2915 is not a
dwarf system in terms of mass, and optical imaging shows that it to be
dominated by an older stellar population for $R > 0.5$ Kpc (MMC94).
These facts suggest that old GCs might have been expected.

Are the number of clusters found anomalous?  To address this we estimate
the total number of clusters in the system.  Unfortunately, our images
sample only a small fraction of the galaxy.  If the GCs are distributed
spherically out to $R_{\rm HI} = 10'$ (where the \HI\ distribution ends;
MCBF96), then we have only surveyed 3.5\%\ of the available area.  Here
we consider two cases for the possible distribution of old clusters.
Case (1) assumes that we are lucky and have managed to observe all the
clusters in NGC\,2915.  While this is unlikely, it provides a strict
lower limit to the cluster formation efficiency.  The more likely case
(2) is that the globular cluster system is like that of more luminous
systems - having a spherically symmetric power-law radial distribution
in number per unit area - $N(R) \propto\ R^\alpha$ \citep{h91}.  We
assumes this extends between $R = 0.5$ Kpc and $R_{\rm HI}$, where the
minimum $R$ insures that the predictions are finite and was chosen to
correspond to the size of the central star forming population (MMC94).
For $-2.5 < \alpha < -0.5$ we estimate that the total number of clusters
in the system is 7 to 12 times higher than found on our images.  For
case (2) we adopt a correction factor of 9.  Hence there are at least 3
GCs in the system (case 1), with a more likely number being $\sim 27$
(case 2).

In principle, we should correct the total GC estimate for the finite
luminosity sampling of the images.  However, SExtractor \citep{ba96}
catalogs of our images show that we can detect extended objects down to
$\vacs \approx 27$ which corresponds to an $M_{V606} \sim -2$ for
sources in NGC2915.  This is well below the peak of the GC luminosity
function $M_{V} \sim -7.5$ (Secker 1992; revised to agree with the {\em
Hipparcos\/} RR Lyrae zero point, e.g. Carretta \etal\
2000)\nocite{s92,cgcfp00}.  Hence we apply no luminosity sampling
correction. The fact that we only have found GCs much brighter than this
peak is somewhat puzzling.

We will consider three measures of cluster formation efficiency.  First,
the specific frequency, $\SN = N_T\, 10^{0.4(M_V + 15)}$, is the $V$
luminosity normalized cluster content \citep{hvdb81}.  \SN\ is typically
around 1 for spiral galaxies and increases towards earlier galaxy types,
with E galaxies having $S_N \sim 4$.  In the center of galaxy clusters
\SN\ ranges from 2.5 to 12.5 \citep{btm97,jpb99}. We consider two
measures of the mass normalized contribution: \etagc\ the number of
clusters per $10^9$ \Msun\ in dynamical mass (Blakeslee \etal\ 1997),
and \epscl, the fractional baryonic (gas and stars) mass in GCs
\citep{McL99}.  As done by \citet{McL99} we assume each cluster has an
average mass $\langle {\cal M}_{\rm cl}\rangle = 2.4 \times 10^5\,
\Msun$ when calculating \epscl.  Blakeslee \etal\ (1997) find an average
$\etagc = 0.7$ with a scatter of 30\%, while \citet{McL99} computes an
average $\epscl = 0.0026$ with a 20\%\ uncertainty.

Table~\ref{t:gcs} tabulates \SN, \etagc\ and \epscl\ for NGC\,2915 under
the above two cases.  Relevant quantities used for these calculations
are given in Table~\ref{t:prop}.  Table~\ref{t:gcs} also lists
literature values and uncertainties of the efficiencies for ``normal''
systems.  The uncertainty of our estimates are large: for a Poissonian
distribution yielding a count of 3, the 95\%\ confidence limits on the
cluster formation efficiencies are 0.3 and 2.3 times the estimated
value.  We find that at a minimum (case 1) \SN\ is close to normal for a
spiral galaxy while while \etagc\ and \epscl\ are low compared to normal
globular cluster systems.  In the more likely case (2), \SN\ is very high
compared to normal gas rich systems.  \etagc\ and \epscl\ are also
somewhat high compared to literature values, but in reasonable agreement
considering the Poissonian uncertainty of our measurements.  While we
can not rule out the possibility that NGC\,2915 is like a normal spiral
galaxy in terms of \SN, it is more likely that it has a high luminosity
normalized cluster content, whereas the mass normalized content is
closer to normal.

\subsection{A missing link in galaxy evolution?}

It is interesting to interpret these results within the framework of
hierarchical evolution through comparison with other systems.
\citet{jpb99} has discussed how the \SN\ of the young Milky Way must
have been fairly high, similar to the values for cluster E galaxies,
after the formation of the Galactic halo but before stellar disk
formation.  The increase in luminosity from later star formation in the
disk would cause a decrease in the Galactic \SN\ to its present low
value, while leaving the number of GCs per unit mass unchanged.  In
NGC\,2915 we have an example of a present-day galaxy with an old stellar
component, including GCs, but a spiral disk that is still mainly in the
form of gas.

In this sense, NGC\,2915, like the centers of rich galaxy clusters, has
an elevated value of \SN\ compared to typical spirals because of a lower
efficiency for converting gas into stars.  Clusters typically have 2--5
times as much mass in gas as in stars \citep{arbvv92}, similar to what
we see in NGC\,2915.  The difference, however, is that the gas in
clusters now resides in the hot intracluster medium, while the gas in
NGC\,2915 still retains the potential to be converted into stars.  To a
large extent, the fate of the gaseous disk (and the future evolution of
\SN\ in NGC\,2915) must depend on the surrounding environment.

Simulations show that when a disk galaxy enters the environment of a
rich galaxy cluster, much of the gas in its disk and halo will be
removed by the combination of tidal and ram pressure stripping
\citep{amb99,bcs02,bcdg01,g03}.  For example, if a system similar to the
Galaxy were to fall into a rich cluster it would have its gas disk
truncated down to a radius of a few kpc within a few tens of Myr by ram
pressure stripping alone (Abadi \etal\ 1999).  Indeed, models of present
day evolution of galaxy clusters invoke mildly truncated star formation
in field galaxies accreting onto the cluster as the cause of the
Butcher-Oemler effect \citep{bnm00,kb01}.  It is likely that in their
{\em early evolution\/} clusters were assembled from building blocks
similar to NGC\,2915 which had their ISM stripped from them by these
mechanisms.  The stripped gas was then virialized to become the cluster
X-ray halo.  In this scenario galaxy clusters have high \SN\ because
they formed out of subclumps which had already efficiently formed star
clusters but which never had the chance to form disks.

In a less hostile environment, a building block similar to NGC\,2915
could go on to form a normal spiral galaxy.  If the \HI\ disk of
NGC\,2915 were to be perturbed enough to efficiently form stars in a
fairly quiescent fashion this would result in an additional $7 \times
10^8 \Msun$ of stars forming with no additional GCs.  Assuming a $M/L_V
\sim 1\, \Msun/L_{V,\odot}$ for the additional stars then the system
would evolve towards $S_N \sim 2$, a fairly normal value for spiral
galaxies, while \etagc\ and \epscl\ remain fixed at their normal values.
The \SN\ in NGC\,2915 is then anomalously high because the formation of
its disk has not proceeded, presumably due to a lack of external
perturbations (MCBF96). If so, then we may expect other galaxies with
high \MHILB\ ratios and extended \HI\ disks to also have a significant
GC population, especially if they have a strong old population.

\begin{figure}
\epsscale{1}
\plotone{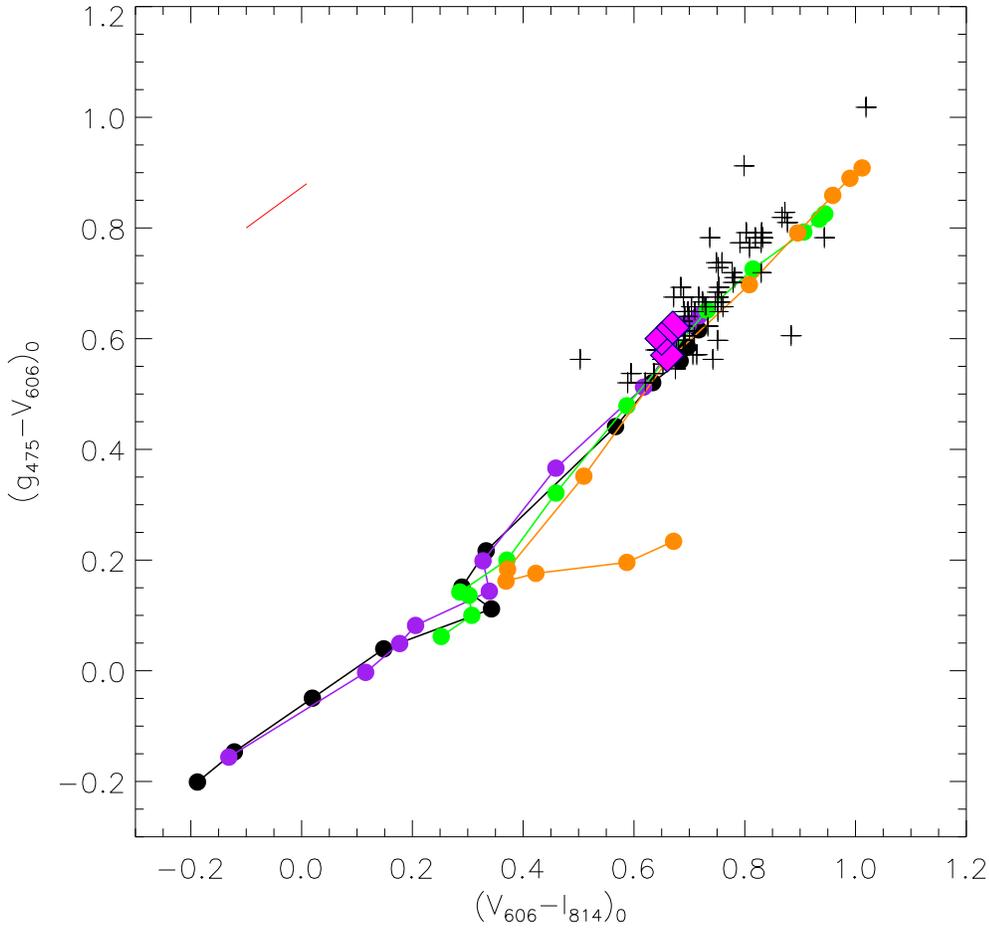}
\caption[]{Color-color plots of the NGC\,2915 globular clusters (pink
filled diamonds) compared to Galactic globular clusters (crosses; from
Harris 1996), and ``simple stellar population'' models from Bruzual \&\
Charlot (2003; colored connected dots).  All data have been corrected
for Galactic extinction. The red line segment shows the $\ebv = 0.1$
reddening vector (Cardelli \etal\ 1999). Population models having ages
of 0.01, 0.02, 0.05, 0.1, 0.2, 0.5, 1.0, 2.0, 5.0, 10, 15, and 20 Gyr
are plotted in four constant metallicity model sequences having $[{\rm
Fe/H}] = $, --2.25 (black); --1.65 (purple); -0.33 (green); and 0.09
(orange). All models employ the standard stellar evolution and Initial
Mass Function prescriptions from Bruzual \&\ Charlot (2003).
\label{f:2cd}}
\end{figure}

While this scenario seems compelling, we caution that it is possible
that NGC\,2915 only superficially resembles the building block we
describe.  We have not proven that the number of clusters is anomalously
high. Furthermore, we can not yet rule out the possibility that the
clusters are of higher metallicity and younger than Galactic GCs.  This
is shown in figure~\ref{f:2cd} where we overplot \citet{bc03} population
models on the two color diagram illustrating the strong age-metallicity
degeneracy for the filters used here.  If the clusters are not old, they
may represent the remnants of a starburst occurring as recently as a few
Gyr ago.  We are undertaking additional observations (imaging and
spectroscopic) to get a more accurate census of star clusters in the
NGC\,2915 system and determine their nature.  The results of these
studies should determine whether NGC\,2915 is just a gas rich galaxy with
a peculiar star formation history, or whether it truly presents us with
a rare local view of how galaxies looked in the epoch of cluster
assembly.

\acknowledgments

ACS was developed under NASA contract NAS 5-32864, and this research has
been supported by NASA grant NAG5-7697.  We thank the technical and
administrative support staff of the ACS science team for their committed
work on the ACS project.

\clearpage

\begin{deluxetable}{lccl}
\tablewidth{0pt}
\tablecaption{NGC\,2915 properties\label{t:prop}}
\tablehead{\colhead{Quantity} &
   \colhead{Value} &
   \colhead{Units} &
   \colhead{Description}
}
\startdata
  \ebv\            & $0.28 \pm 0.04$      & mag   & foreground extinction \\
  $D$              & $4.1 \pm 0.3$        & Mpc   & Distance \\
  ${\cal M}_g$     & $7.4 \times 10^8$    & \Msun & ISM mass \\
  ${\cal M}_\star$ & $3.2 \times 10^8$    & \Msun & Mass in stars \\
  ${\cal M}_T$     & $2.1 \times 10^{10}$ & \Msun & Total dynamical mass \\
  $M_V$            & --16.42              & mag   & Absolute mag $V$ band \\
  $L_B$            & $3.4 \times 10^8$    & $L_{B,\odot}$ & $B$ band luminosity \\
  \MTLB\           & 62                   & solar & Mass to light ratio \\
\enddata
\end{deluxetable}

\begin{deluxetable}{l l l c c c c c c c}
\tablewidth{0pt}
\tablecaption{Cluster properties\label{t:clust}}
\tablehead{  \colhead{ID} &
  \colhead{RA\tablenotemark{a}} &
  \colhead{Dec\tablenotemark{a}} &
  \colhead{$R$} &
  \colhead{$M_{V606}$} &
  \colhead{\small \GVacs} &
  \colhead{\small \VIacs} &
  \colhead{\rhalf} &
  \colhead{$\epsilon$} \\
  \colhead{ } &
  \colhead{\footnotesize (J2000)} &
  \colhead{\footnotesize (J2000)} &
  \colhead{\footnotesize (kpc)} & 
  \colhead{ } &
  \colhead{ } &
  \colhead{ } &
  \colhead{(pc)} &
  \colhead{ } \\
  }
\startdata
G1 & 09 25 56.273 & -76 35 14.53 & 3.0 & -9.82 & 0.62 & 0.66 & 8.8 & 0.11 \\
G2 & 09 25 27.445 & -76 36 31.55 & 3.3 & -9.04 & 0.57 & 0.65 & 4.6 & 0.16 \\
G3 & 09 25 41.954 & -76 36 33.70 & 2.4 & -8.92 & 0.59 & 0.64 & 5.9 & 0.13 \\
\enddata
\tablenotetext{a}{The absolute accuracy of the positions $\sim
0{\farcs}2$ in each coordinate is set by the astrometric calibration 
which employed 86 stars in the HST Guide Star
Catalog version 2 (http://www-gsss.stsci.edu/gsc/gsc2/GSC2home.htm).  
The relative accuracy of the positions is
$< 0{\farcs}05$ in each coordinate.}
\end{deluxetable}

\begin{deluxetable}{l c c c }
\tablewidth{0pt}
\tablecaption{Globular Cluster System properties\label{t:gcs}}
\tablehead{\colhead{Quantity} &
  \colhead{Case (1)}     &
  \colhead{Case (2)}     &
  \colhead{Literature}}
\startdata
\SN     & 0.81    & 7.3    & 0--12.5             \\
\etagc  & 0.14    & 1.3    & $0.7 \pm 0.2$       \\
\epscl  & 0.00067 & 0.0061 & $0.0026 \pm 0.0005$ \\
\enddata
\end{deluxetable}

\clearpage

\end{document}